# Conformal Three-Dimensional Interphase of Li Metal Anode Revealed by Low Dose Cryo-Electron Microscopy

Bing Han[†,1], Xiangyan Li[†,1,4], Shuang Bai[2], Yucheng Zou[1], Bingyu Lu[3], Minghao Zhang[3], Xiaomin Ma[5], Zhi Chang[6], Ying Shirley Meng[2,3]*, Meng Gu[1]*

[1]Department of Materials Science and Engineering, Southern University of Science and Technology, Shenzhen 518055, China

[2]Materials Science and Engineering, University of California San Diego, La Jolla, 92093, CA, USA

[3]Department of Nano Engineering, University of California San Diego, La Jolla, 92093, CA, USA

[4]Academy for Advanced Interdisciplinary Studies, Southern University of Science and Technology, Shenzhen, 518055, China

[5]Cryo-TEM Center, Southern University of Science and Technology, Shenzhen 518055, China

[6]Energy Technology Research Institute, National Institute of Advanced Industrial Science and Technology (AIST), Tsukuba 305-8568, Japan

[†]These authors contributed equally to this work.

*Correspondence to: gum@sustech.edu.cn ; shirleymeng@ucsd.edu

**Abstract:** Using cryogenic transmission electron microscopy, we revealed three dimensional (3D) structural details of the electrochemically plated lithium (Li) flakes and their solid electrolyte interphase (SEI), including the composite SEI skin-layer and SEI fossil pieces buried inside the Li matrix. As the SEI skin-layer is largely comprised of nanocrystalline LiF and $Li_2O$ in amorphous polymeric matrix, when complete Li stripping occurs, the compromised SEI three-dimensional framework buckles, forming nanoscale bends and wrinkles. We showed that the flexibility and resilience of the SEI skin-layer plays a vital role in preserving an intact SEI 3D framework after Li stripping. The intact SEI network enables the nucleation and growth of the newly plated Li inside the previously formed SEI network in the subsequent cycles, preventing additional large amount of SEI formation between newly plated Li metal and the electrolyte. In addition, cells cycled under the accurately controlled uniaxial pressure can further enhance the repeated utilization of the SEI framework and improve the coulombic efficiency (CE) by up to 97%, demonstrating an effective strategy of reducing the formation of additional SEI and inactive "dead" Li. The identification of such flexible and porous 3D SEI framework clarifies the working mechanism of SEI in lithium metal anode for batteries. The insights provided in this work will inspire researchers to design more functional artificial 3D SEI on other metal anodes to improve rechargeable metal battery with long cycle life.

**One Sentence Summary:** With applied uniaxial stack pressure, a stable and porous 3D SEI network allows for fast $Li^+$ transport and the nucleation of newly plated Li metal without the formation of large quantities of additional SEI.

**Main Text:**

SEI has been heavily researched in the past few decades, due to its importance in extending Li-ion battery life and capacity retention[1-3]. The dynamic growth and fracturing of the SEI during battery cycling are closely correlated with the reactions between the electrolyte solutions and the electrode surfaces[4-8]. Ideally, the SEI should be formed during the initial cycles and then serves as a barrier between the bulk Li and the electrolyte solution, in favor of preventing continuous consumption of the Li ions and electrolyte solution in the subsequent cycles. Consequently, many researchers have tried to optimize the composition of electrolyte in order to form a stable and uniform thin passivating SEI on the anodes. However, previous studies on the SEI are heavily reliant on spectroscopic techniques, such as X-ray photoelectron spectroscopy (XPS), tip enhanced Raman[9], nuclear magnetic resonance (NMR)[10], and other advanced electroanalytical methods.[11-14] For example, Peled *et al.* depict a mosaic SEI structure, where the inorganic SEI components (nanosized $Li_2O$, $LiF$, $Li_2CO_3$, etc.) seem to be assembled randomly on the anode surface. To determine whether these SEI arrangements are randomly formed or driven by a yet-unseen mechanism, analytical techniques with high spatial resolution are needed to accurately probe the identity and distributions of these nano phases. However, due to the nature that the SEI samples are extremely sensitive to air and probing source such as electron beams, accurate measurement of the composition, component distribution, and morphology of the SEI at the atomic scale is rarely reported, not to mention the 3D information. To gain more insights into the failure mechanism of Li metal anodes, we need a clearer map of the SEI evolution at different charge and discharge states, which will help us to further understand the role of SEI layers and their relation to the performance of the Li metal anode.

Because of the fragility of liquid-solid interfaces in Li metal anodes, the best option for atomic-scale SEI characterization is cryogenic aberration-corrected transmission electron microscopy (cryo-TEM).[3,15-18] Using this method, Wang *et al.* revealed the nucleation process of Li metal, and demonstrated that glassy Li metal is more favorable in electrochemical reversibility;[17] Cui *et al.* reported the distinct mosaic and multilayer types of SEI that forms on the Li-metal anode after cycling with pure EC/DEC and FEC-modified EC/DEC electrolytes, respectively, and demonstrated that FEC-modification dramatically improves battery performance.[19,20] Cryo-electron energy loss spectroscopy mapping (EELS) has also revealed the possible formation of LiH phase in plated Li metal, which may reduce the cycling capability of Li metal anode.[18] The SEI skin-layer generally contains two important types of components, inorganic crystalline grains and an organic, amorphous polymeric matrix[21]. The distribution of inorganic components inside the polymeric matrix plays a critical role in dictating the ionic and electronic conductivity[22] and mechanical stability of the SEI, and thus the cyclability of the Li metal anode. The ideal SEI skin-layer keeps its integrity by resiliently accommodating the volume changes of the anode materials during many cycles of Li plating and stripping. The mechano-chemical properties, resistance, ionic conductivity, and recyclibity of the SEI urgently need precise characterization at high resolution in real space[19,23-25]. However, only very few studies are available in the literature to unveil the atomic structure, leaving a largely unknown domain for research. The morphology and structure of the SEI after Li deposition and stripping need to be further studied at the atomic scale and three dimensions using cryo-TEM in order to accurately correlate its structure-property with Li metal cyclability.

...

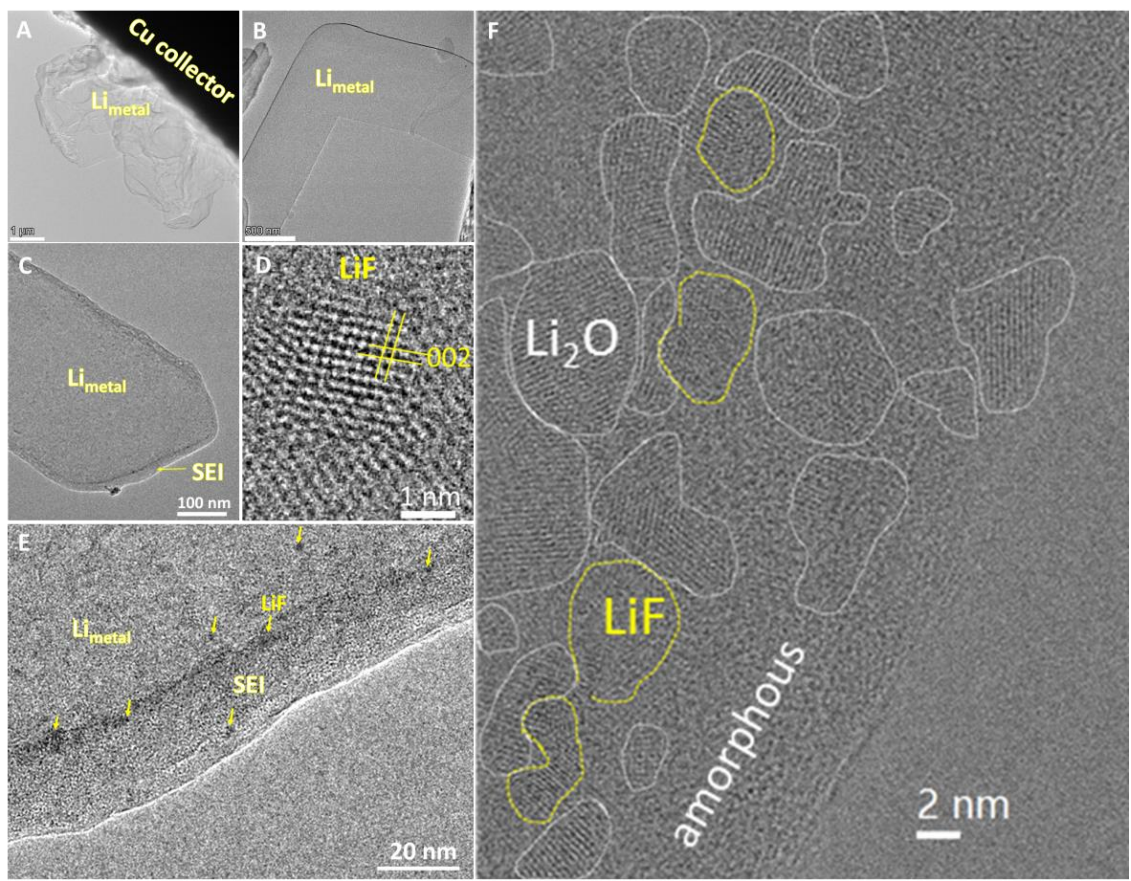

**Figure 1. Cryo-TEM analysis of the plated Li flake and SEI formed using FEC additive after the 1st plating process. A-C**. Low-magnification cryo-TEM images showing plated Li metal and its SEI; **D**. HRTEM showing the LiF nanocrystal inside the SEI; **E**. Magnified local region showing the dark LiF nanocrystals (indicated by yellow arrows) inside the SEI; **F**. Distribution map of different phases in the SEI skin-layer. (The HRTEM images in panel D&F are acquired with electron dose rate ~8 e Å$^{-2}$ s$^{-1}$ for ~10s).

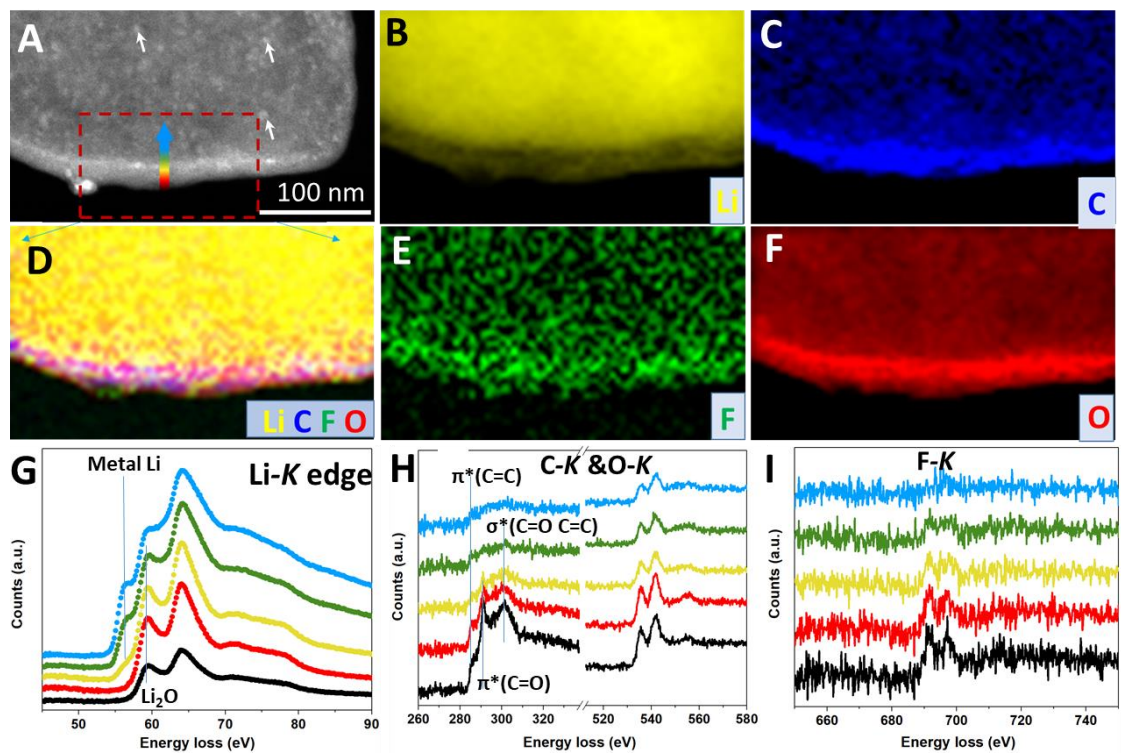

**Figure 2. STEM EELS analysis of the plated Li metal and its SEI.** (**A**) STEM image showing the region of the EELS map (red rectangle) and line profile (arrow); (**B-F**) EELS elemental maps of Li, C, composite, F, and O; **G**. Li *K* edge; **H**. C *K* and O *K* edges; **I**. F *K* edge from the line profile along the arrow from the surface to the inside of the Li metal. Panel B-F shares the same scale bar with panel A.

The LiF component in the SEI in FEC-modified EC/DEC is believed to be the primary factor of enhanced performance. However, its presence is heavily debated in the field. Li *et al.* did not find LiF in their study[20], while others reports the identification and important roles of LiF in stabilizing the SEI.[26-28] We believe that electron dosage plays vital roles in accurately imaging the native state of the SEI on Li metal. Therefore, we developed a low-dose technique to allow acquisition of HRTEM images with only 80 e Å$^{-2}$, compared to the 30000 e Å$^{-2}$ chosen in Li *et al.* study[20]. We focused our study on the higher-performing FEC-modified EC/DEC electrolyte solution. We characterized the same sample from the micron-scale down to the atomic-scale, revealing that the as-plated Li metal exhibits flake morphology ranging from a few hundred nanometers to 10 microns in **Figure 1A-C**. The FEC additive induces the formation of a uniformly thin SEI layer (~25 nm) on the Li metal (**Figure 1B**) with the LiF nanocrystals distributed in the inner layer of the SEI skin (**Figure 1C-E**). The HRTEM in **Figure 1D** clearly shows the {002} set of planes of LiF. At the same time, large quantities of $Li_2O$ were found in the surficial SEI skin as shown in **Figure 1F**. The nanocrystalline LiF and $Li_2O$ islands are distributed randomly inside the amorphous polymeric matrix, which forms the general morphology of the insulating SEI skin. In contrast, the SEI formed in pure EC/DEC is very non-uniform with large variation in thickness, and are composed of mostly $Li_2O$, and $Li_2CO_3$ components as shown in **Figure S1**. The $Li_2CO_3$ component is unstable against electrolyte, and will decompose to form gaseous species.[21] Therefore, the addition of FEC can induce a uniformly thin SEI layer with more stable LiF inorganic crystals and minimize unstable $Li_2CO_3$ component.[21]

The combination of our cryogenic scanning transmission electron microscopy (cryo-STEM) and cryo-EELS results offers unprecedented insight into the chemical composition of the SEI skin-layer after just the 1st plating cycle. As shown by the large-scale cryo-STEM in **Figure 2A**, there are a high density of bright clusters (indicated by white arrows) against the dark Li matrix, demonstrating local enrichment of impurities. The cryo-EELS elemental maps in **Figure 2B-F** also detect Li, oxygen (O), fluorine (F), and carbon (C) signals inside the Li matrix. We speculate that the presence of the impurities (including fluorine, carbon, and possibly oxygen) is one of the important factors that lead to the amorphization of some local regions of the plated Li because the local fluctuation and enrichment of these impurities leads to the formation of nanocrystalline $Li_2O$, and LiF (which may break SEI skin-layers into pieces) randomly buried inside the Li matrix during the electrochemical cycling. These EELS elemental maps agree well with the HRTEM shown in **Figure 1F**, which shows the presence of $Li_2O$ and LiF nano-islands inside the plated Li matrix. The EELS line profiles indicate that, when scanning from the surface to the bulk, the intensity of the peaks of the elements (F, O, and C) has an obvious high-to-low trend (**Figure 2H-I**). The simultaneous increase of metallic Li and the decrease of the elements such as F, O, and C strongly suggests that SEI elements have been doped into the plated Li metal matrix. The same cryo-EELS fine structure analysis also shows that the amorphous polymeric matrix contains large amounts of carbonate groups (–O–(C=O)–O–) (**Figure 2H**). Furthermore, previous literature also indicates that SEI formed using FEC-containing electrolyte contains mostly cross-linked (ethylene oxide) (PEO) and aliphatic chains along with carbonate and carboxylate species[10], which is consistent with our EELS analysis. The SEI skin-layer contains oxidized Li, while the inner Li bulk exhibit more metallic fine structure of Li *K* (**Figure 2G-H**), and the SEI skin-layer has a strong F signal, while the inner matrix has a relatively weak F signal (**Figure 2I**). Therefore, the elemental content between the plated Li metal and SEI skin are not as distinct as we generally speculated.[4]

The SEI skin acts as a protective layer for the plated Li metal, preventing further reaction between the electrolyte and Li metal. The estimated electron tunneling distances may be as short as 2-3 nm for perfectly compacted inorganic SEI components (such as $Li_2O$, $LiF$, or $Li_2CO_3$) as predicted by Density Functional Theory (DFT) calculations.[29,30] However, even though the as-formed SEI is a composite of both inorganic crystals and organic components and may contain structural defects, a 20-30 nm thick SEI skin should largely prevent electron leakage and protect the deposited Li metal.

In order to evaluate the structural robustness of the SEI skin-layer, we probed the morphology of the SEI after the first stripping cycle (**Figure 3**). The filled SEI skin-layer with large flakes of the plated Li metal inside (**Figure 3A**) contrasts sharply with the empty, deflated, and empty SEI husks in (**Figure 3B**). Interestingly, the process of Li stripping caused the empty SEI husks to shrink without complete collapse, which indicates that the mechanical strength of the SEI allows it to survive large volume changes, probably because of the flexibility of the organic/polymeric matrix in the SEI skin-layer. While the study by Li *et al.* demonstrated the importance of having an electrolyte solution (FEC additive) that can lead to an organic/polymeric matrix in the SEI skin-layer, our characterization explains *why* those components are so critical. The Selected Area Electron Diffraction (SEAD) pattern further indicates that $Li_2O$ is the dominating inorganic component of the empty SEI husks in **Figure 3C**. The High Angle Annular Dark Field (HAADF) STEM images and elemental maps in **Figure 3D** shows the content of F, C, O, and Li in the empty SEI husks. These results are in good agreement with the previously identified $Li_2O$ and $LiF$ nanocrystalline regions in the SEI.

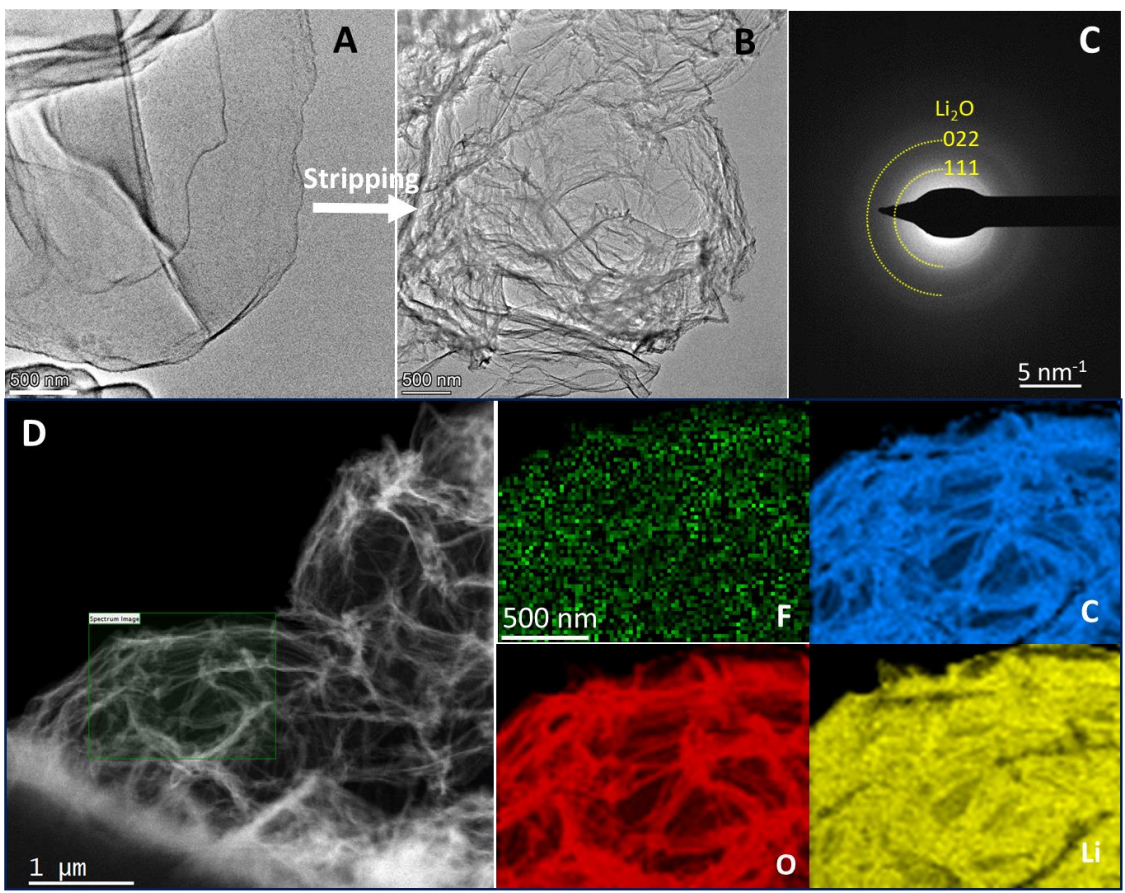

**Figure 3.** (**A**) Cryo-TEM image of the as-plated Li; (**B**) morphology of the remaining empty husks after first stripping cycle; (**C**) Electron diffraction of the remaining empty husks after Li stripping; (**D**) cryo-STEM image and EELS elemental maps (green: fluorine; blue: carbon; red: oxygen; yellow: Li) of the empty husks.

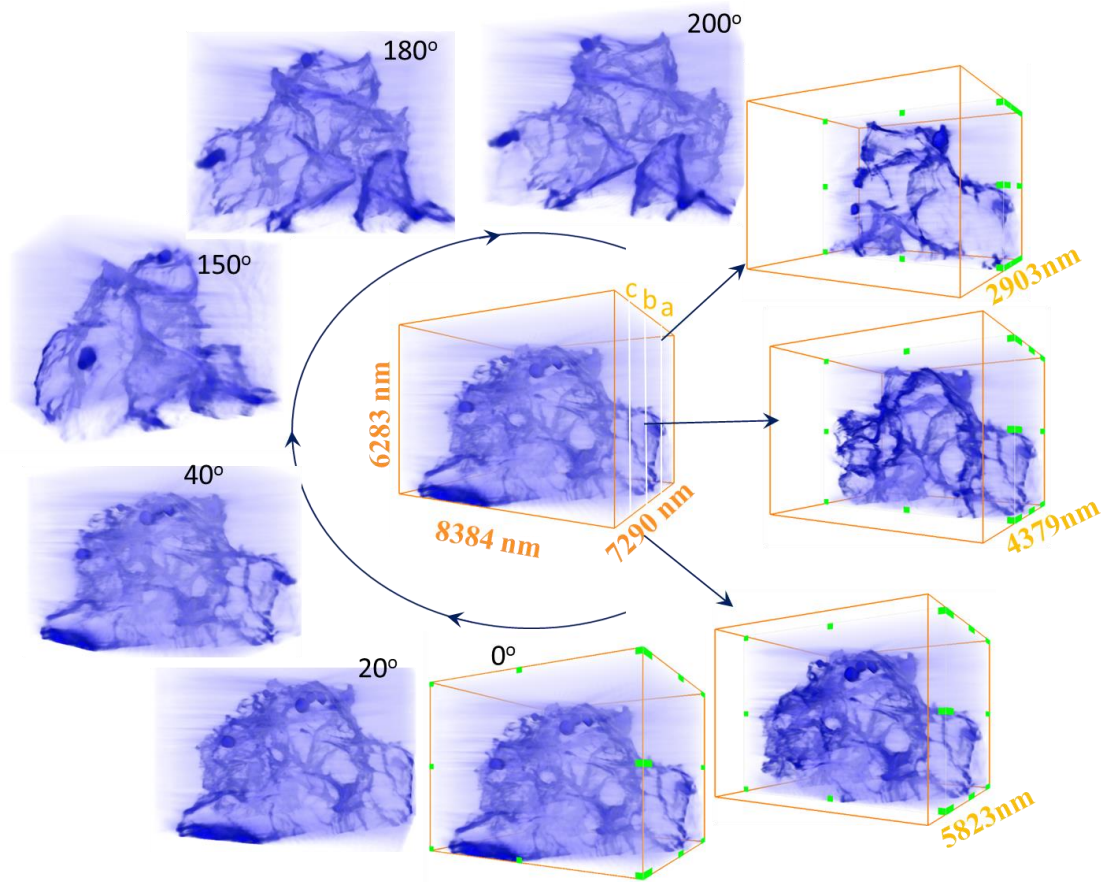

**Figure 4. STEM tomography reconstructed the 3D image of the SEI husk after Li stripping, illustrating the hollow, crumpled SEI structure.** The left images (0°-200°) show the 3D SEI husks viewed from different angles; while the right images (boxed in yellow) show cross-sections the SEI husks at the position of 'a' 2903 nm, 'b' 4379 nm, and 'c' 5823 nm (marked by the white lines in the central image). (For the complete tilt series, 0-360° rotational view, and cross-section slicing, please refer to **Movies S1-S3**).

In order to visualize the 3D structure of the SEI network more clearly, we performed a cryo-STEM tomography tilt series in a 0-100° degree range (**Figure S2** and **Movie S1**) to fully reveal the empty SEI husk network. We also displayed the reconstructed 3D SEI structure from 0-360° angles (**Movie S2**) and in a cross-section slicing video (**Movie S3**). **Figure 4** illustrates the empty 3D husks at selected viewing angles and cross-section slices. As clearly shown by the tilt series and reconstructed 3D videos, the buckled SEI husks remain intact but porous after Li stripping (see also **Figure S3** for smaller but deflated SEI husk). The flexibility of the amorphous SEI matrix allows the SEI to fold, and bend during the stripping, which preserve the intact frameworks of the SEI. The cross-linked polymeric phase in the SEI skin-layer keeps the SEI framework intact during Li plating by resiliently cushioning the volume changes during stripping and platting, a flexibility that is largely correlated with both good capacity retention and high CE.[31,32]

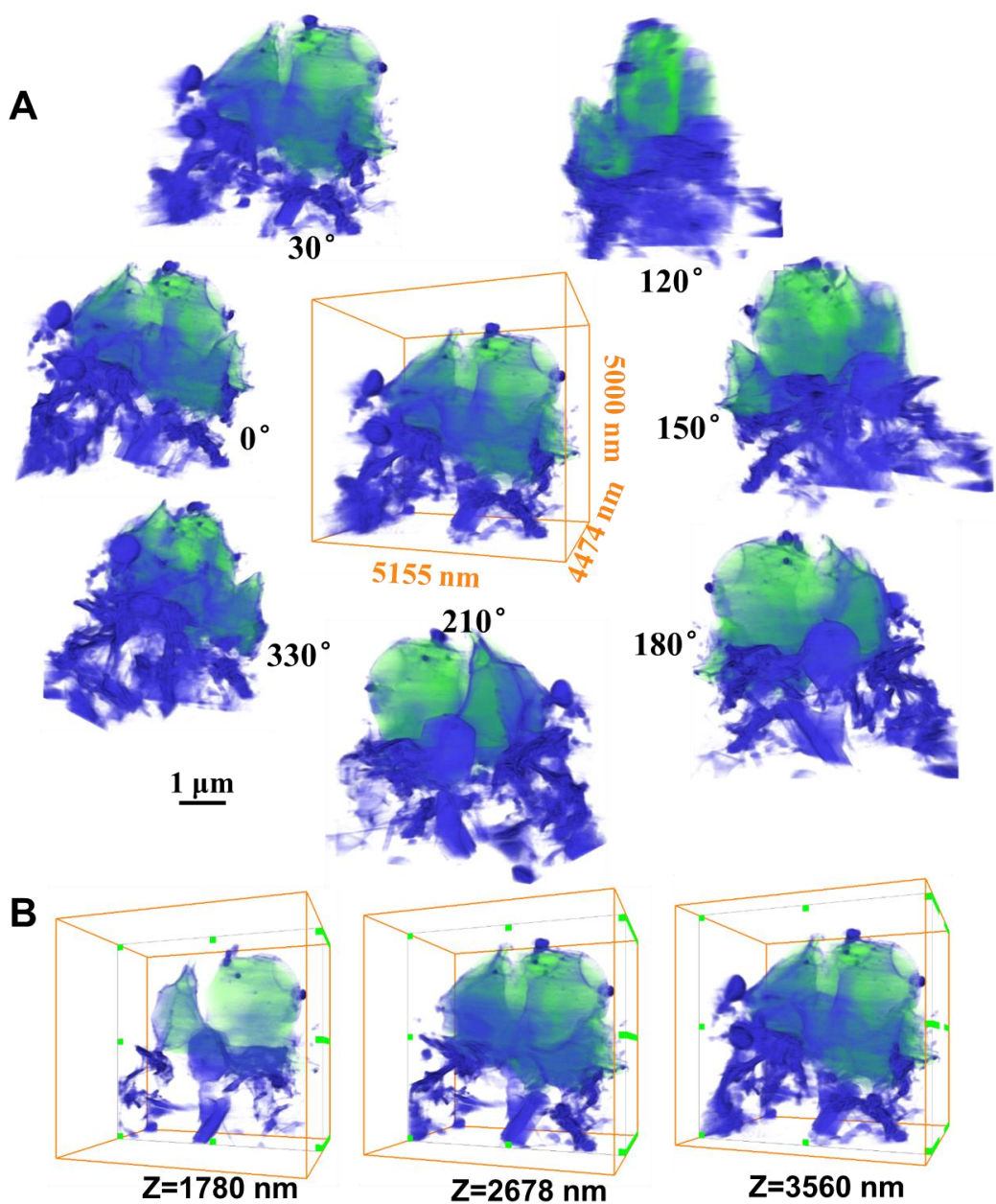

**Figure 5.** STEM tomography reconstructed 3D image of the dead Li (highlighted in green color) after Li stripping. A. 0°-360° view showing the dead Li particle viewed from different angles; B. cross-section views at different positions of Z-axis (1780 nm, 2678 nm, and 3560 nm). (For better visualization, please refer to **Movies S4-S5**).

In addition to identifying the empty SEI husks after Li stripping, 3D STEM tomography also helps to accurately locate the presence of dead Li (highlighted in green color) in **Figure 5**. The 0-360º view in **Figure 5a** and **Movie S4** clearly identifies that the dead Li particle (~4μm) sits on top of the empty SEI husks and its electron conduction path is cut off from the bottom part of the Li metal or the current collector. As a result, the Li atoms in this particle can no longer lose electrons to current collector and become free $Li^+$ ions. Furthermore, the cross-sectional views in **Figure 5b** and **Movie S5** illustrate the morphology variations in the dead Li particle with changing cross-section slices and its connections with SEI husks below. In addition, **Figure S4** in the supporting information shows additional dead Li metal after just one stripping cycle due to likely non-uniform local stripping rates during the stripping cycle. The observation of the dead Li formation in 3D allows us to see the origin of dead Li, which is the primary reason why batteries lose capacity and deteriorates during cycling.[3]

We further note that the empty SEI husks form a perfect three-dimensional network for the newly plated Li in the following cycle. To better understand how the Li re-fills the SEI husks, we analyzed a new Li electrode after two lithiation cycles. As anticipated, the empty husks are reflated by the newly plated Li (**Figure 6A-B**). The HAADF STEM image and elemental maps in **Figure 6B-C** also show the presence of O, F, and some carbon species inside the Li metal matrix.[21] The high contrast of O inside the Li matrix clearly indicates the formation $Li_2O$ and the network-like distribution of $Li_2O$. Interestingly, we used multiple linear least square (MLLS) fitting to deconvolute the Li *K* edge to show the presence of metallic Li and oxidized Li in the EELS maps (**Figure 6C**). The SEI skin-layer is rich in $Li_2O$, while the bulk Li metal primarily contains metallic Li, with only a small amount of $Li_2O$. Again, we see that the 3D SEI network allows Li to penetrate through the SEI, nucleate, and grow without forming new SEI.

The addition of FEC into the EC/DEC obviously enhances the electrochemical cycling performance of Li metal anode to a large extent.[11,20,33,34] Our understanding of the SEI's 3D structure inspired us to further tailor the setup of the battery cell to optimize cycling performance. As observed in our experiments, the size distribution of the empty SEI-husks is quite large, ranging from a few hundred nanometers to about ten micrometers. Such large quantities of high-density, insulating SEI husks on the surface of the Li metal inevitably leads to increased impedance of the cells after the first cycle. In addition, the formation of dead Li is clearly observed in **Figure 5**. Therefore, to facilitate the Li transport and re-deposition of Li into the empty SEI husks and minimize the formation of dead Li, we designed a split cell with precisely controlled uniaxial stack pressure (350 kPa) to increase the reusability of the SEI husks as shown in **Figure 7A**.[35] An intimate contact with the current collector ensures Li to re-fill into the SEI husks. Our optimized uniaxial stack pressure greatly improved CE (**Figure 7B**) of 96% at the 1st cycle and 97% at the 6th cycle, vastly outperforming the same EC/DEC-FEC cell without pressure (CE ~87%-92%).

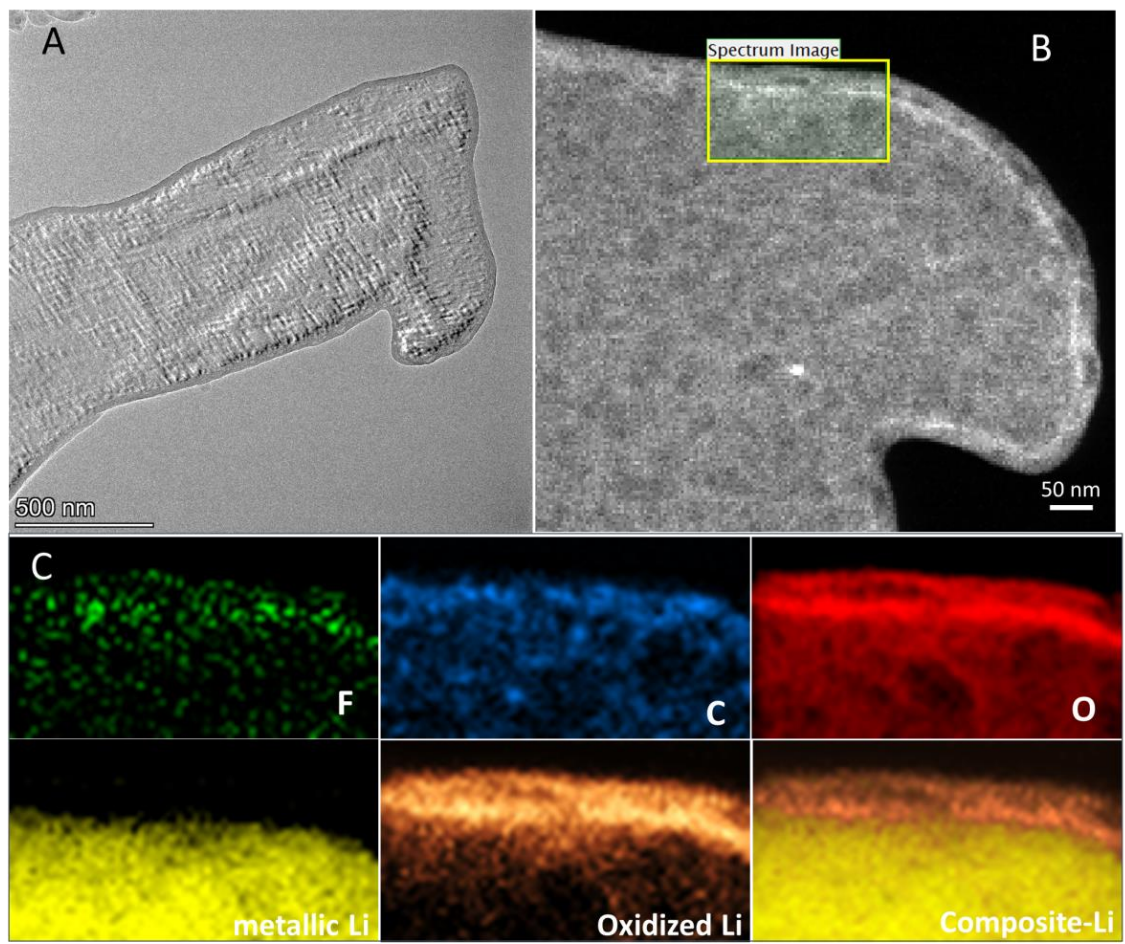

**Figure. 6**. (**A**) cryo-TEM and (**B**) cryo-STEM image of the 2nd round of plated Li metal; (**C**) elemental maps of F, C, O, metallic Li, oxidized Li, and composite Li map in the rectangle spectrum image region in panel **B**.

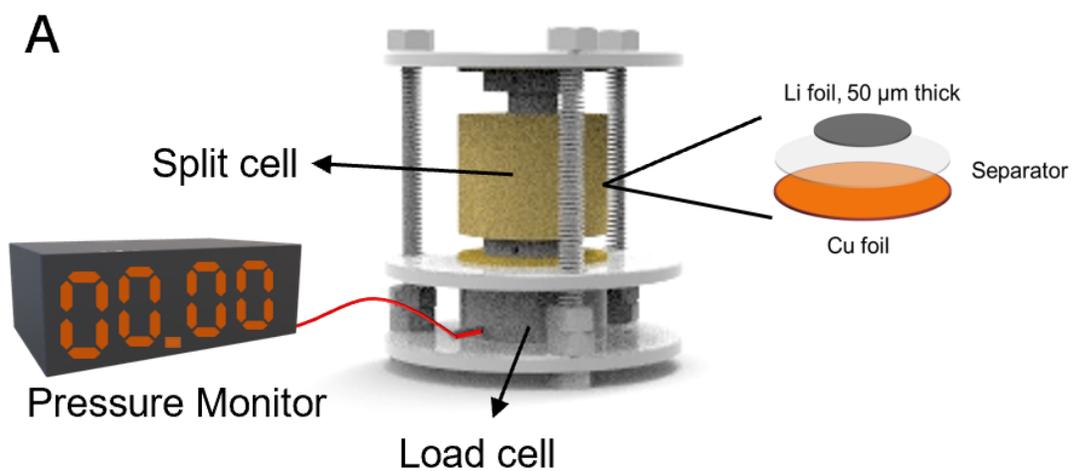

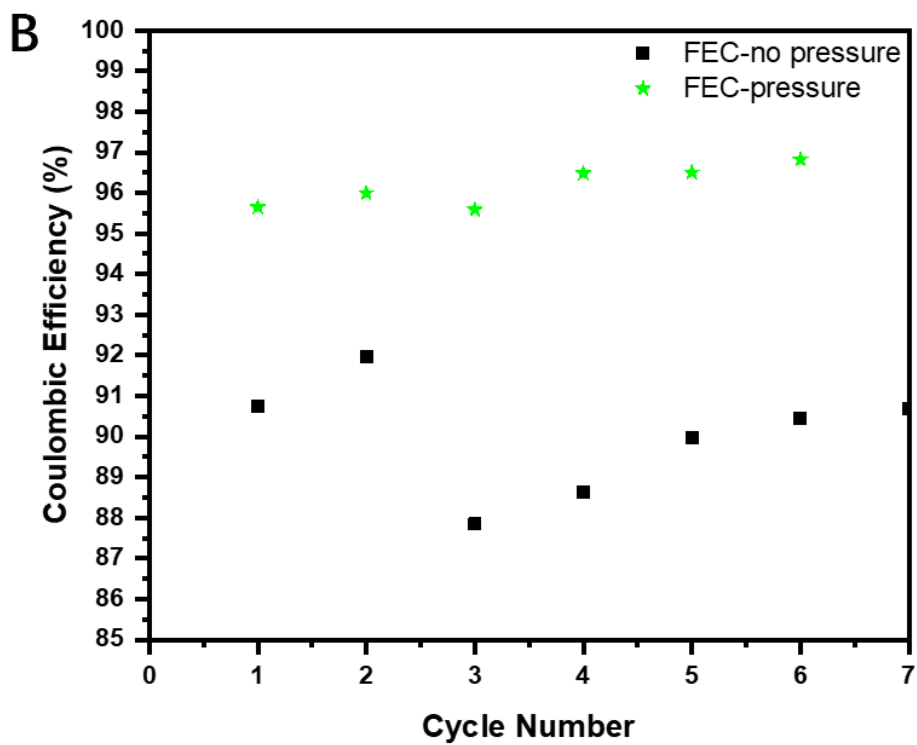

**Figure 7. (A)** schematic showing the pressure-cell with precisely controlled uniaxial stack pressure (350 kPa) applied to the battery during cycling; **(B)** cycling CE of the pressure-cell compared with no pressure using 1M $LiPF_6$ electrolyte solutions with EC:DEC:FEC (20:70:10 by vol%).

**Conclusions**

Cryo-STEM tomography is an effective tool for visualizing the atomic structure and 3D architecture of the SEI at different charge and discharge states. Using cryo-TEM, we demonstrated that a good (FEC-additive) electrolyte solution induces the formation of a 3D SEI network with a uniform organic/polymeric outer layer and a LiF-containing inner layer. Because the SEI inherits flexible, stretchable property from the organic SEI layer and a robust structure from the inner, inorganic SEI layer, the 3D SEI husk crumples but does not collapse during delithiation. In fact, the SEI husk facilitates $Li^+$ transport and nucleation and growth of Li in the next lithiation cycle because the flexible empty husk morphology allows Li to refill the SEI husks and limit the formation of new SEI layers. This characterization explains well-documented observation that the CE improves after the first cycle, during which the 3D SEI network forms.

In addition, we put a Li||Cu cell under accurately controlled stack pressure and demonstrated that a closer and intimate contact with the current collector allows the SEI husks to be more repeatedly used in the following cycles and further improved CE by up to 5~9% in the cycling. Our study offers a successful example of the kind of improvement that this atomic-scale and 3D characterization can induce, and provides exciting new analytical method for future SEI and other beam-sensitive materials.

**Experimental:**

We assembled CR2032-type coin cells in an Ar-filled glove box using Li metal as counter electrode. We used 1M LiPF$_6$ electrolyte solutions with EC:DEC:FEC (20:70:10 by vol%). A Cu-TEM grid was intentionally put on to the Cu surface in the working electrode. The Li plating and stripping cycles are conducted with a constant current mode of 0.5 mA/cm$^2$ to a capacity of 0.5mAh/cm$^2$. After cycling, we opened up the coin cells and take the Cu-TEM grid out to transfer into the TEM after rinsing with DMC to remove residue salts and electrolytes. We used a Titan Krios with Cs-aberration corrector operating at 300 kV. All TEM images are acquired at 77K with ultra-low dose of electrons (~7 e·Å$^{-2}$·s$^{-1}$ × 10 s) with direct detect device (Falcon 3). The low-magnification STEM used a low current of 11pA and the dosage for STEM is around 200 e/nm$^2$. The cryo-EELS maps were acquired using 11pA current with 0.1 s dwell time at each pixel and the EELS spectra are summed over a few pixels to get better signal to noise ratio.

Li-Cu split cell test: A customized split cell with a load cell was used to precisely control the uniaxial stack pressure (350 kPa) applied to the battery during cycling. The split cell consists of two parts: two titanium plungers and one polyether ether ketone (PEEK) die mold. The Cu‖Li cells were assembled in an Ar-filled glovebox by sandwiching the Li metal foil (50 µm thick), Celgard separator, and the cleaned Cu foil between the two titanium plungers inside the PEEK die mold. Only minimum amount of electrolyte (~5 µL) was added to the Cu‖Li cells to wet the separator.

**Supplemental information:**

Supplemental Information includes four figures and five movies, and can be found with this article online at http://dx.doi.org/xxxxxxxx.

Submitted Manuscript: Confidential

**Acknowledgments:** This work was supported by Guangdong Innovative and Entrepreneurial Research Team Program (2016ZT06N500), Shenzhen Peacock Plan (KQTD2016022620054656), Shenzhen DRC project [2018]1433, Shenzhen Clean Energy Research Institute (No. CERI-KY-2019-003). This work was performed at the Pico and Cryo-TEM Center at SUSTech Core Research Facility that receives support from Presidential fund and Development and Reform Commission of Shenzhen Municipality. Y.S.M. acknowledges funding support from Zable Endowed Chair of Energy Technology and the Sustainable Power & Energy Center of UC San Diego. S. B. is grateful for the financial support from Materials Science and Engineering Program of UC San Diego.

**Author contributions:** M.G., Y. S. M. designed and supervised the experiment. B.H., Z.Z., Y. Z. carried out the cryo-TEM and battery testing; X. L. did the tomography reconstruction, S. B., M. Z. M.G., B.H., J. L., Z.Z. analyzed the data; all authors discussed and edited the paper.